# Semi-supervised Variational Temporal Convolutional Network for IoT Communication Multi-anomaly Detection

Yan xu* Yongliang Cheng

*Abstract*—The consumer Internet of Things (IoT) have developed in recent years. Mass IoT devices are constructed to build a huge communications network. But these devices are insecure in reality, it means that the communications network are exposed by the attacker. Moreover, the IoT communication network also faces with variety of sudden errors. Therefore, it easily leads to that is vulnerable with the threat of attacker and system failure. The severe situation of IoT communication network motivates the development of new techniques to automatically detect multi-anomaly. In this paper, we propose SS-VTCN, a semi-supervised network for IoT multiple anomaly detection that works well effectively for IoT communication network. SS-VTCN is designed to capture the normal patterns of the IoT traffic data based on the distribution whether it is labeled or not by learning their representations with key techniques such as Variational Autoencoders and Temporal Convolutional Network. This network can use the encode data to predict preliminary result, and reconstruct input data to determine anomalies by the representations. Extensive evaluation experiments based on a benchmark dataset and a real consumer smart home dataset demonstrate that SS-VTCN is more suitable than supervised and unsupervised method with better performance when compared other state-of-art semi-supervised method.

## I. Introduction

Internet of Things (IoT) [1] systems are playing an important and vital role in the human smart dwelling environment. Meanwhile, it motivates the rapid increasement of smart services that can improve the entire Quality of Service (QoS) in smart home [2]. However, many of these IoT devices are fundamentally insecure. There are many latent bugs such as Open telnet ports, outdated system firmware, and unencrypted transmission of sensitive data, which may rise the security risk by hacker malicious attack or some potential system failures in IoT communication [3]. In order to reduce the cost of failures and attacks, anomalies in IoT communication must be discovered accurately in time.

While anomaly detection for extensive time series scenarios have been solved reasonably well by neural networks [4], the IoT communication anomaly detection have been studied based on fundamental machine learning methods. These existing anomalies in IoT communication are prevalent and intricate, which are caused by multifarious reasons and have unique solving strategies corresponding to anomaly type. Moreover, the capture communication data distribution of label influences the detection model construction in a great extent. Nowadays, the supervised anomaly detection achieve mature development with the help of Recurrent Neural Network (RNN) and so on [5]. However, the generation of communication data is fast and large. It means that the large amount communication data is difficult to label. Hence, some unsupervised methods like Autoencoder [6] and Generative Adversarial Networks (GAN) [7] are devoted to solve anomaly detection problems without label. However, the unsupervised methods cannot clearly define the anomaly category and disturb the security policy selection without labels. And the researchers can easily distinguish the anomaly and label a part of communication data in general. Therefore, the semi-supervised methods is a suitable compromise selection based on the realistic condition [8].

Anomaly detection has been an active research topic in data mining with applications in graph, log messages, time series, etc. This paper further employs semi-supervised deep learning method on account of early anomaly detection research in IoT environment. In this scenario, apart from the label distribution problem, the large amount of communication records are captured in chronological order [9]. There are temporal relationships between each record, which we cannot easily make out. And the characteristic of IoT communication data, tender temporal relationships are fully considered by us in this paper.

Firstly, we defined the IoT communication data as a kind of time series data. At this point, Recurrent Neural Network (RNN) [5] is an effective deep learning neural network, which has been explored for time series data. Especially, Long Short-Term Memory (LSTM) [10], an improved neural network of RNN, performs better than other variants in majority sequence problems. However, Temporal Convolutional Network (TCN) [11] is a specially constructed convolution network, which is proposed to solve time series problems by retaining more valid long-term memory. And the constructed network is simpler and more accurate than RNNs in majority sequence problems. The extensive experiments indicates TCN is able to achieve better performance in many sequence problems. Hence, we adopt TCN as a discriminative model to distinguish the types of anomalies. Then, Variational Autoencoder (VAE) is a deep Bayesian network [12], which contains two random variables $x$ and $z$. $p(x)$ is called empirical distribution and $p(z)$ is called prior distribution. Meanwhile, VAE can generate new samples by computing conditional distribution $p\theta(x|z)$. We employ VAE as a generative model to generate new samples based on original samples. At last, we combine VAE and TCN into a more complex network named SS-VTCN to achieve classify and generating function.

Moreover, we design a novel loss function to constraint the SS-VTCN training process based on part of labeled data and large amount unlabeled data. It is a key point to construct this semi-supervised model. And the contributions of this paper are listed as follows.

1) SS-VTCN, an innovative semi-supervised anomaly detection model combines the discriminative model TCN and the generative model VAE. SS-VTCN can effectively detect multi-anomaly by learning representation of time series and robust probability distribution. Hence, SS-VTCN applies to solve the anomaly detection problem in the real IoT communication scenario.

2) We design a novel loss function for SS-VTCN which considers the effect of TCN and VAE. The discriminative model TCN is adopted for labeled data which is aimed to search the difference of anomaly type. Moreover, the generative model VAE can deal with explicit temporal dependence among stochastic variables to learn robust representations of labeled and unlabeled data. Hence, SS-VTCN can get the anomaly type prediction which is restrained by the novel loss function.

3) We present a series of rectification principles to decrease the rate that the preliminary prediction result is incorrect. In fact, the preliminary prediction is trained by a small amount of labeled data. Hence, the rectification principles entirely consider the flaw of preliminary prediction and the latent representations ability to improve the performance.

4) We experiments on real IoT communication dataset for multi-anomaly detection by our proposed method and other representative method. Meanwhile, the experiment results demonstrate the effectiveness of our approach.

The rest of this paper is organized as follows. In section II, we review the related work. In section III, we state the preliminaries of the problem setting and overall structure. In section IV, we present our proposed method for multi-anomaly detection in IoT communication. In section V, we show the experimental results of different datasets and methods. Moreover, we demonstrate the proposed model is effective and efficient for solving the problem. Finally, we conclude our work of this paper in section VI.

## II. RELATED WORK

Anomaly detection is a general problem, that there are different methods available, including supervised [13], unsupervised [14], distance-based [15] and density-based [16]. However, the rapid development of IoT accompanies a great challenge of communication anomaly detection. Cauteruccio et al. [17] proposed a new methodological framework that can make future investigations in this research field easier, coherent, and uniform. However, this is a framework for anomaly detection in IoT rather than a specific method for solving communication anomaly detection. Cook et al. [18] provides a background on the challenges which may be encountered when applying anomaly detection techniques to IoT data, with examples of applications for the IoT anomaly detection taken from the literature. Among these challenges, the anomaly detection in smart home is very severe. Hence, Fahad et al. [19] proposed a model that can recognize approach the activities performed in a smart home, and separates the normal from the anomalous activities. Meanwhile, Tang et al. [20] proposed an ensemble model anomaly detection method which can target the data anomalies present in general smart Internet of Things (IoT) devices, allowing for easy detection of anomalous events based on data. Moreover, Xu et al. [21] concentrate on the anomaly detection problem of smart home services and distinguish different anomalies of communication. However, these approaches need a large amount normal activities as training data and should pay a heavy price.

Besides supervised models, the research of unsupervised anomaly detection is abundant and obtain a lot of achievement. Pajouh et al. [22] proposed a novel model for intrusion detection based on two-layer dimension reduction and two-tier classification module, designed to detect malicious activities such as User to Root (U2R) and Remote to Local (R2L) attacks. However, the performance can be improved to a certain extent by deep learning method. Lin et al. [23] proposed a VAE-LSTM hybrid model as an unsupervised approach for anomaly detection in time series. Kim et al. [24] proposed a payload-based abnormal behavior detection method, named the autoencoder-based payload anomaly detection (APAD). However, these methods is hard to detect different kind anomaly by unsupervised learning way, which the background is not based on IoT communication.

Actually, there are some researches, which base on semi-supervised learning and accord with the IoT communication. Ruff et al. [25] proposed Deep SAD, an end-to-end deep methodology for general semi-supervised anomaly detection and achieve excellent performance. Miao et al. [26] proposed a distributed online OCSVM for anomaly detection over networks and get a decentralized cost function in IoT communication. Cheng et al. [14] [27] proposed a semi-supervised Hierarchical Stacking Temporal Convolutional Network (HS-TCN), which is a semi-supervised model for anomaly detection in IoT communication. However, these above approaches have a serious problem, that they are not suitable to employ for muti-anomaly detection problem. Therefore, it is necessary to design a semi-supervised neural network for muti-anomaly detection, which is effective and efficient in IoT communication.

## III. PRELIMINARIES

In this section, we first present the problem statement of anomaly detection in Internet of Things (IoT) communication. Then, we show the overall structure of our model named SS-VTCN. Finally, we introduce Temporal Convolutional Network (TCN) [11] and Variational Autoencoder (VAE) [12], which are the key components of our model.

### A. Problem Statement

The communication data contain successive observations which are usually collected at equal-space timestamps in IoT. We focus on the time series of communication, defined as $X = \{X_1, X_2, \cdots, X_t\}$. The $t$ is the length of $X$ correspond to the communication chronological order. Since this anomaly detection problem, the data is time series data, which is not belong to discrete data. Hence, we set the previous part of original dataset as training data and the rest is test data. In this way, we obey the principle, that we train the history data to predict the future communication record whether is anomaly. As for each communication records, we use $X_t = \{x_1, x_2, \cdots, x_u\}$ to denote at the numerical order $t$. The $x_u$ means the feature. $u$ is the dimension size of features [28]. There is no difference in feature representation of communication record whether the records have label.

As for semi-supervised model, the dataset of training is complex, which contain labeled and unlabeled datasets. In general, the previous records are suitable to label with less cost and enough time. We defined the labeled dataset as $D^l$. The records of $D^l$ are denoted as $[X_t^l, Y_t]$ at the numerical order $t$.

The unlabeled dataset are defined as $D^u$ and the records are denoted as $[X_t^u]$. For the anomaly detection in IoT communication, the objective is to confirm the communication record whether is belonged to anomaly or not. Meanwhile, we extensive use to fit the training data for understanding test data in constructing time series model. It is important that both training and test are not allow the future records to influence the previous. That means the sequence of $\{X_1, X_2, \cdots, X_t\}$ not only $X_t$ is used to detect the anomaly result of $X_{t+1}$. The goal of this paper is to build a model $f: x \rightarrow y$, so that it can be further used to recognize anomalies [25].

### B. Overall Structure

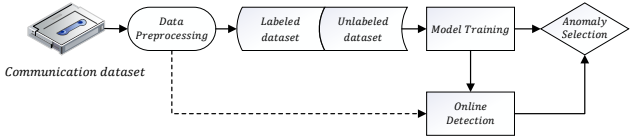

Figure 1. The process of SS-VTCN

As shown in Figure 1, the overall structure of IoT communication multi-anomaly detection contain two parts: offline training and online detection [4]. Offline training does not have real-time requirement. This part aims to constantly train multi-anomaly detection model for updating online detection. Online detection guards the IoT communication secure by real-time detecting anomalies all the time. Data Preprocessing is important which can obtain the training dataset by labeling communication and following chronological order. The training dataset is divide into two datasets which are labeled and unlabeled. Because of the hysteretic nature on labeling records, the time of labeled records is previous compared to unlabeled records. Meanwhile, the above modules are shared by both offline training and online detection.

After preprocessing and dataset dividing, the communication records are send to Model Training module. The major structure of Model Training module is SS-VTCN, which is a semi-supervised model and combines TCN [11] and VAE [12]. The TCN can process long input sequences as a whole, instead of sequentially such as LSTM [10], RNN [5] and so on. Although there are a large portion of unlabeled training records, it has no negative influence for searching time dependence and storing long effective memory. Because of the existence of anomaly type in labeled data, TCN can maximumly utilize these labels for multi-anomaly detection. Hence, our model solves the problem that unsupervised model is difficult to determine the exact anomaly type. Moreover, the regularization of all data make the model have good properties of generative process by VAE. Then, the mature model by Model Training module is offer to Online Detection. Finally, both online and offline model can effectively detect the anomalies and update the corresponding model by Anomaly Result.

### C. TCN and VAE

Recurrent Neural Networks (RNN) [5] and its advanced and creative variants such as LSTM have achieved a great development for solving sequence modeling problems. However, the appearance of Temporal Convolutional Network (TCN) [11] brings an enormous challenge for RNNs. Because TCN has architecture simplicity, autoregressive prediction, and very long memory without gating mechanisms. Besides, abundant experiments demonstrate that TCN performs better than RNNs in dealing with most sequence modeling problems.

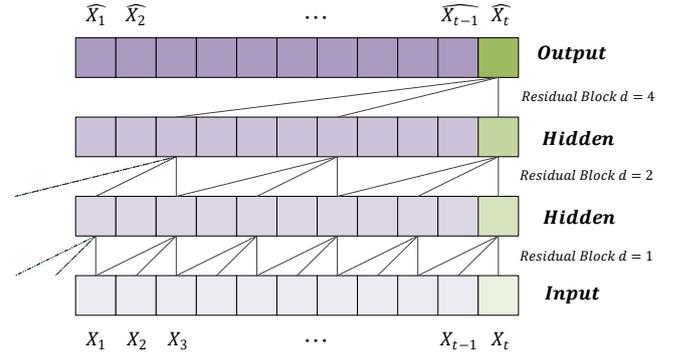

Figure 2. The structure of TCN

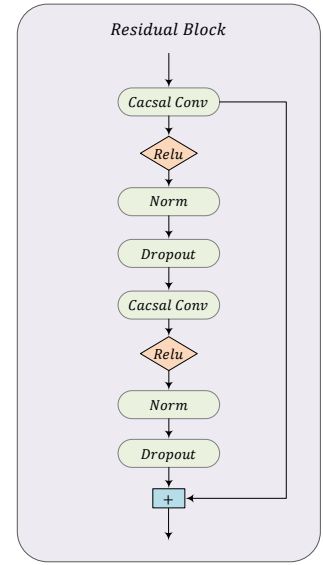

Figure 3. The structure of Residual Block

As shown in Figure 2, TCN can achieve the network produces an output of the same length as the input. Meanwhile, TCN would have no leakage from the future into the past. TCN mainly uses the 1D fully-convolutional network (FCN) [29] and causal convolutions to achieve the above performance. Moreover, the architecture of TCN can be formulated as:

$$TCN = 1D\ FCN + causal\ convolutions \quad (1)$$

The causal convolution can only look back the history in the depth of the network. Moreover, the application of dilated convolutions makes a larger receptive field. The input $X_t$ and the filter $f: \{0, \cdots, w-1\}$ constitute the dilated convolutions operation of $F$ on the sequence $s$ can be formulated as:

$$F(s) = \sum_{i=0}^{w-1} f(t) \cdot X_t^{s-d \cdot i} \quad (2)$$

The $d$ is dilation factor and the $w$ is filter size. Meanwhile, the $s - d \cdot i$ means the direction of history. Moreover, TCN uses a residual block to contain a series of transformations $F$ as shown in Figure 3. The $\widehat{X_t}$ is the output of residual block which is added to the $x_t$ and formulated as:

$$\widehat{X_t} = Activation(X_t + G(X_t)) \quad (3)$$

There is no information missing for the TCN architecture. Besides, TCN has enough input length flexibility which proves that convolution is also effective and efficient for sequence modeling problems. More contents about TCN can be found in [].

VAE has been successfully applied to anomaly detection by generating data, which belongs to deep Bayesian model [4] [12]. The performance of VAE is usually better than common Autoencoder(AE) which appear overfitting phenomenon easily. VAE represents a high-dimensional input $\widehat{X_t}$ to a latent representation $Z_t$ with a reduced dimension, and then reconstructs $\widehat{X_t}$ by $Z_t$. The $p_\theta(Z_t)$ is prior to $Z_t$, while $\widehat{X_t}$ is sampled from posterior distributions $p_\theta(\widehat{X_t}|Z_t)$. However, it is difficult to compute $p_\theta(\widehat{X_t}|Z_t)$. Hence, VAE designs an inference network $q_\emptyset(Z_t|\widehat{X_t})$ for approximating $p_\theta(\widehat{X_t}|Z_t)$. The architecture of VAE is shown in Figure 4.

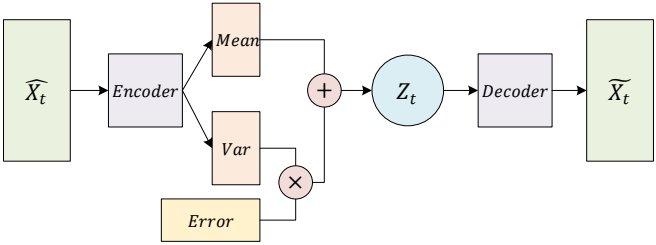

Figure 4. The structure of VAE

Because of the ingenious architecture design, VAE can get the latent variable $Z_t$ randomly sampled from a prior data distribution. VAE comprises inference network $q_\emptyset(Z_t|\widehat{X_t})$ as an encoder for encoding input data $\widehat{X_t}$ into latent space $Z$.

Actually, VAE encoder would export two vectors named $Mean$ and $Var$. Meanwhile, there is a loss variable named $Error$. VAE uses these variable to calculate $Z_t$, which is different from other Autoencoder. Finally, the decoder $p_\theta(\widehat{X_t}|Z_t)$ uses the corresponding $Z_t$ for approximating generation data $\widetilde{X_t}$.

IV. PROPOSED METHOD

In this section, we first present the semi-supervised network architecture of SS-VTCN, followed by offline model training, online detection and anomaly selection.

A. Network Architecture

The network of SS-VTCN is mainly divided into the following parts. First, it uses Temporal Convolutional Network (TCN) [11] to capture complex temporal dependence between the continuous communication records during long time span. Then, we apply Variational Autoencoder (VAE) [12], a popular and effective generative model for representation learning. Although TCN and VAE own these advantages for training, it is difficult to combine there excellent neural network, which functions are entirely different. Especially, it is necessary and delicate to design the semi-supervised model based on a large part of records without labels. Actually, our proposed network SS-VTCN, a semi-supervised model has the advantages both of TCN and VAE. Meanwhile, SS-VTCN can utilize less labels to fit all records and minimize the reconstruction loss to predict more accurately.

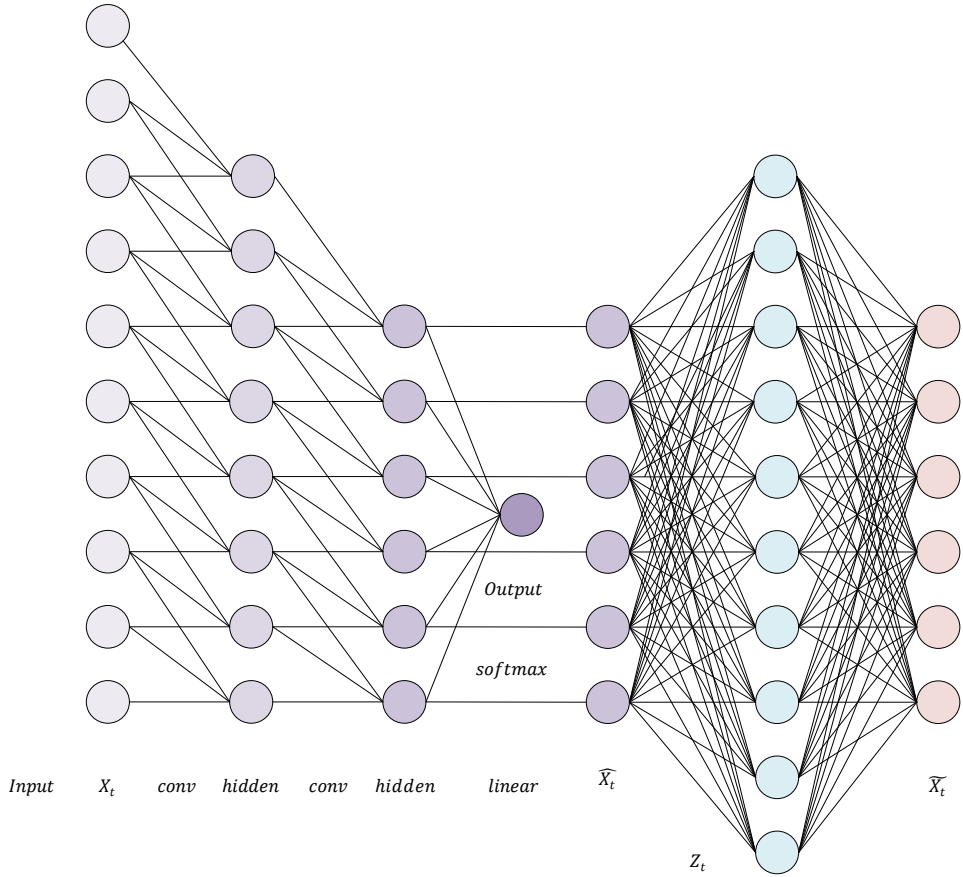

Figure 5. The structure of SS-VTCN

The overall SS-VTCN network architecture is shown as Figure 5, which is composed of TCN and improved VAE [4]. The capture process of TCN is semi-supervised, which uses part of communication records with labels for these records and other records without labels. The adoption of TCN can effectively learn time-series distributions in latent stochastic space. In addition, this part can acquire training detection results by the labeled records for calculating loss function. On the other hand, the improved VAE is different from the ordinary VAE as shown in Figure 4, which aims to map high dimension to low dimension for encoding and encoding part to high dimension. Because of the TCN training result $\widehat{X_t}$ contains abundant temporal dependence as a low dimension, we improve VAE for aim to map the lower dimension $\widehat{X_t}$ from TCN to higher dimension $Z_t$. Similarly, the reconstruction error between $\widehat{X_t}$ and $\widetilde{X_t}$ is also used for calculating loss function. Moreover, $Z_t$ is also used for approximating generation data $\widetilde{X_t}$.

### B. Model Training

The TCN and VAE in SS-VTCN are trained simultaneously by tuning the network parameters [30]. We can train our model straightforwardly. Because of only less of records have labels, it is difficult to minimize the loss for semi-supervised model SS-VTCN rather than other supervised models. The loss function $\mathcal{L}_{SS-VTCN}$ is composed of $\mathcal{L}_T$ and $\mathcal{L}_V$, which is formulated as:

$$\mathcal{L}_{SS-VTCN} = \mathcal{L}_T + \mathcal{L}_V \quad (4)$$

$\mathcal{L}_T$ represents the loss of SS-VTCN left part as shown in Figure 5, which is mainly concerned by TCN. On account of the semi-supervised learning way, we improve the general TCN loss function named $\mathcal{L}_T$. As for $\mathcal{L}_V$, it represents the loss of SS-VTCN right part as shown in Figure 5, which is the reconstruction loss of VAE. In addition, the output of VAE based on minimizing $\mathcal{L}_V$ is aimed to predict more accurately [4].

In fact, $\mathcal{L}_T$ is more complex than $\mathcal{L}_V$. For the semi-supervised model design, $\mathcal{L}_T$ is the most important. It also decides that model utilizes less records with labels and improves the performance based on these priori knowledge. Because of SS-VTCN is a semi-supervised model, there are two training dataset respectively named unlabeled dataset $D^u$ and labeled dataset $D^l$. Meanwhile, the length of $D^u$ is $N^u$ and the length of $D^l$ is $N^l$. For each anomaly detection problem, $M$ means the number of anomaly category. In addition, $y_{ic}$ and $p_{ic}$ respectively mean the class indicator variable and the prediction probability of class $c$ for labeled dataset, which is same as $y_{jc}$ and $p_{jc}$ for unlabeled dataset. However, $y_{jc}$ means the class indicator variable of the last step prediction rather than the real label like $y_{ic}$. Actually, the initial prediction are defined as class 0. For the class indicator variable $y$, it can be formulated as:

$$y = \begin{cases} 0 & output \neq label \\ 1 & output = label \end{cases} \quad (5)$$

The $output$ means the prediction of training dataset as show in Figure 5. In summary, $\mathcal{L}_T$ can be formulated as:

$$\mathcal{L}_T = \frac{1}{N^l+N^u}(\sum_{i=1}^{N^l} -\sum_{c=1}^{M} y_{ic}\log(p_{ic}) + \sum_{j=1}^{N^u} -\sum_{c=1}^{M} y_{jc} * p_{jc}\log(p_{jc})) \quad (6)$$

As for the other part of SS-VTCN, it is aimed to minimize the reconstructing loss. We make the loss function of VAE accomplish $\mathcal{L}_V$, which is formulated as:

$$\mathcal{L}_V = -\frac{1}{N^l+N^u}\sum_{i=1}^{N^l+N^u} E_{q_\phi(Z_t|\widehat{X_t})}[\log(p_\theta(\widehat{X_t}|Z_t))] + KL(q_\phi(Z_t|\widehat{X_t})\|p(Z_t)) \quad (6)$$

The $KL(\cdot\|\cdot)$ is the Kullback-Leibler (KL) divergence [31] and $p(Z_t)$ is the prior distribution. KL divergence ensures that VAE can distil latent space information for reconstructing input data. Meanwhile, the prior distribution can promote the posterior distribution to be close to prior distribution.

### C. Online Detection and Anomaly Selection

Now we can determine whether each communication record at a moment is anomalous or not using SS-VTCN after model training. Note that the input of SS-VTCN is a communication sequence data of a moment. Thus, we take the sequence $X_t$ to TCN training for searching temporal dependence and have the $output$ as the preliminary prediction result by SoftMax function. Meanwhile, the last linear information $\widehat{X_t}$ as the input of next model part. Then, $\widehat{X_t}$ is extended to the dimension of previous communication sequence, which is named as $Z_t$. In addition, $\widetilde{X_t}$ is the reconstruction data of SS-VTCN. As suggested by [32], this reconstruction between $\widehat{X_t}$ and $\widetilde{X_t}$ of communication sequence is conditional probability, which is named as $p^v$. The evaluation of $p^v$ is formulated by:

$$p^v = \log(p_\theta((\widehat{X_t}|\widetilde{X_t})) + \log(p_\theta((X_t|Z_t)) \quad (7)$$

Meanwhile, $p^v$ is applicated to rectify the preliminary prediction result. Specifically, the reconstruction probability $p^v$ is used to offer reference for the last anomaly delection.as the anomaly score in our model. Because of the majority of communication records is normal, it means that high $p^v$ means the original communication records can be well reconstructed. On the other hand, the lower of $p^v$, the less likely the observation can be reconstructed. Meanwhile, this communication record is more likely to be anomalous. In general, we can define an anomaly threshold. If the reconstruction probability $p^v$ is lower than the anomaly threshold, then the corresponding communication record $X_t$ can be marked as anomaly. Otherwise, $X_t$ is normal.

However, there are several kinds of anomaly. It increase the difficulty of ascertaining the anomaly threshold value. It is important to achieve anomaly selection by determining the anomaly threshold. Actually, SS-VTCN is a semi-supervised model and there are some communication records with labels, which belongs to training data. During the training process, we also can get the reconstruction probabilities of these communication records. Besides, the labels of these communication records is easily acquired. Hence, we ascertain the boundary range whether the communication record is normal or abnormal by comparing the label to reconstruction probability. Finally, we design the rectification principles about anomaly selection. Next, we specifically describe the specific rectification principles.

As for the multi-anomaly detection problem in IoT communication, it makes the rectification process more complex. There are five combination possibilities of the $output$ and reconstruction probability $p^v$. The first possibility is that the result of $output$ is normal and $p^v$ is same. The second possibility is that the result of $output$ is normal and $p^v$ is abnormal. The third possibility is that the result of $output$ is abnormal and $p^v$ points to the same kind anomaly. The fourth possibility is that the result of $output$ is abnormal and $p^v$ does not point to the same kind anomaly. The last possibility is that the result of $output$ is abnormal and $p^v$ is normal. The first and third possibilities mean that the final result is uncontroversial. As for the second and last possibilities, the result of $p^v$ is selected to rectify the preliminary prediction result of $output$. The rest fourth possibility means that the preliminary prediction result of $output$ are not rectified. Moreover, the specific rectification principles are presented in Table I. In conclusion, our proposed online detection and anomaly selection parts are applicable to semi-supervised anomaly detection[4].

TABLE I
RECTIFICATION PRINCIPLES

| $output$ | $p^v$ | Status | Rectification |
|---|---|---|---|
| Normal | Normal | Same | No |
| Normal | Abnormal | Different | Yes |
| Abnormal | Abnormal | Same | No |
| Abnormal | Abnormal | Different | No |
| Abnormal | Normal | Different | Yes |

## V. PERFORMANCE EVALUATION

In this section, we first describe the basic condition of IoT communication datasets and experiment setting. Then, we introduce the performance metrics and comparison models. Finally, we conduct many experiments to show the effectiveness of the proposed model in detail.

### A. Datasets and setting

To demonstrate the effectiveness of our proposed model SS-VTCN, we conduct our experiment on a real IoT communication training dataset. These a training dataset is captured from the Distributed Smart Space Orchestration System (DS2OS) [9] [33]. This system can manage many networked and smart devices by solving the large amount and heterogeneity of the devices. It is a new management middleware design with abundant efficient potential. Meanwhile, this dataset is named as DS2OS-B. Moreover, the dataset DS2OS-B is sampling from the original dataset by a novel concept drift adaptive sampling method [21]. In addition, there is a common test dataset named DS2OS-T, which is aimed to examine generalization performance of detection models based the training dataset. An comprehensive summary of these datasets are given in Table 2.

Table II. Evaluation Datasets

| Dataset name | No. of All | No. of Normal | No. of Dos | No. of Malicious | No. of Spying |
|---|---|---|---|---|---|
| DS2OS-B | 16,000 | 8,000 | 4,315 | 2,689 | 996 |
| DS2OS-T | 156222 | 150,053 | 3,780 | 2,317 | 72 |

As for dataset DS2OS-B, it is sampling from the original dataset which has 16,000 communication records. The content of dataset is communication information between smart devices in DS2OS environment. Meanwhile, this DS2OS environment contains several types devices such as light controller, thermometer, movement sensors, washing machines and so on. The dimension of this communication record is 11, while each dimension represents respective meaning, such as resources ID number, resource address, resource type, timestamp, and so on. Moreover, there are four classes of the communication record label, which respectively are normal, Dos attack, malicious and spying. It is important that there are 50% abnormal records. It means that DS2OS-B can be approximated as a balanced dataset. Although the rate of each kind anomaly is different, the rate gap is slight which may not lead to long tail effect.

For the last dataset DS2OS-T, it is a test dataset which is separated from the original dataset except the dataset before sampling. It's worth noting that the communication data in IoT environment is following the chronological order. It means that these datasets are belong to streaming dataset. It is important that we must obey the rule, which is that model only can train the past communication record to detect the future communication record. Hence, the communication record of test dataset is fresher than the communication records of training dataset.

In addition, the proposed model and compared model are developed and coded in Pytorch [34], which is a deep learning framework based on Python released by Facebook. The experiment environment is running on a personal computer, which configuration is Intel i5-8400, 8 GB RAM, NVIDIA GeForce GTX 980. The personal computer operating system is Windows 10. The hyper parameters of each neural network respectively are hidden layer $|l| = 8$, hidden units per layer $|c| = 8$, learning rate $lr = 0.005$ and epoch number $e = 8$. Moreover, the parameter optimization is set as Adam. Besides, the ratio setting of labeled data includes five layers, which are 10%, 20%, 30%, 40% and 50%.

### B. Comparison of models and evaluation criteria

To prove the effectiveness of SS-VTCN, the comparison models are introduced in next part.

- LSTM: LSTM [10] network has shown its ability in handling sequential modeling problems and has superiority in encoding long-term dependencies. Additionally, LSTM is applied to full labeled data for anomaly detection and classification as a supervised learning way.

- TCN: TCN [11] network is a supervised learning way that demonstrates that TCN has better performance in many sequential problems. TCN utilizes the advantages of convolution structure and learns long time dependence, which is suitable for solving some problems likes anomaly detection.

- VLSTM: VLSTM combines VAE [12] and LSTM [10], which is similar as the network of SS-VTCN. However, there is different that the TCN part is changed by LSTM. In addition, VLSTM is a supervised model, which is designed to train labeled data.
- VTCN: The network of VTCN is similar as SS-VTCN. However, VTCN is a supervised model, which is different from SS-VTCN.
- SS-VLSTM: SS-VLSTM is a semi-supervised model. The design philosophy is the same as SS-VTCN. However, we applicate LTSM to exchange TCN.
- SS-WVTCN: SS-WVTCN is a semi-supervised model. In addition, SS-WVTCN is a unique version, that SS-WVTCN does not employ the rectification principles. It can be regarded as a weak variety of SS-VTCN.

As for these above mentioned comparison models, the front 4 models are supervised model. Moreover, the LSTM and TCN are the typical neural networks, which prove the effectiveness and achieve excellent performance. However, VLSTM and VTCN follow the network of proposed model and compare to typical neural network, which is aimed to demonstrate the proposed network design is effective. In other hand, these supervised models cannot apply unlabeled data. Actually, they train the same size labeled data when compared to these semi-supervised models. The last two models are semi-supervised models, which also are the varieties of SS-VTCN. However, there are obvious differences that the basic neural network and principle component are not same as SS-VTCN. In next part, there are sufficient experiment results.

In addition, we employ $F1$ score to evaluate the performance of proposed model SS-VTCN and these comparison models [35]. However, we aimed to solve a multi-anomaly detection problem. We need use each kind anomaly and the normal criteria to evaluate global performance. Hence, we adopt $F1$ score of each class to evaluate performance giving comprehensive consideration to $Precison$ and $Recall$. $Precison$ and $Recall$ are respectively defined as: $Precision = \frac{True\ Positive}{True\ positive + False\ Positive}$, $Recall = \frac{True\ Positive}{True\ positive + False\ negative}$. Moreover, $F1$ score is defined as: $F1 = \frac{2*Precision*Recall}{Precision+Recall}$.

## C. Experiments

This section shows the experiment results and analyses the experiments about the proposed SS-VTCN model and other models. Firstly, we not only adopt $F1$ score of the normal and each kind anomaly as experiment criteria, but also employ the average $F1$ score of each class as an aggregative indicator. The average $F1$ score is named as $Avg$. Then, the rate of labeled data size may have an obvious influence of performance. Hence, we use five layers to evaluate the performance of semi-supervised and supervised models. Finally, Table 3 shows the performance of these semi-supervised and supervised models at DS2OS-B by training dataset DS2OS-T. From the experiment results, we can conclude that:

- There is an obvious improvement that semi-supervised models have better performance than supervised models. As shown in Table 3, there is about 10%-50% improvement in $F1$ score whichever the labeled rate is in training dataset. It means that three semi-supervised models effectively train the unlabeled data and have excellent influence in the detection performance.
- SS-VTCN and VTCN have the same network which likes to SS-VLSTM and VLSTM. However, the semi-supervised models of the same network, SS-VTCN and SS-VLSTM is superior to VTCN and VLSTM. As shown in Table 3, there is about 10%-30% improvement in $F1$ score at each labeled data rate. In addition, the performance of the models based TCN is better when compared to the models which is based LSTM. In conclusion, the network of SS-VTCN is suitable for semi-supervised muti-anomaly detection.
- SS-WVTCN is a different from SS-VTCN, which lacks rectification principles. As shown in Table 3, the performance of SS-VTCN is better than SS-WVTCN. Hence, the rectification principles effectively improve the performance for the muti-anomaly detection.
- Comparing to the models which is based LSTM, almost all models based TCN get better performance. As shown in Table 3, the models based TCN prove the robust effectiveness and stability, which demonstrate that TCN is more suitable than LSTM in applicating for muti-anomaly detection.
- By the reason of the growth of the labeled data rate is expensive and difficult. It's found that the participation of labeled data influence the detection performance to a great extent. In general, the rate of labeled data is larger, the performance is better. However, the 40% of labeled data is relatively reasonably as shown in Table 3. It means that design of proposed model fully considers the unlabeled data and avoids the overfitting by large rate of labeled data.
- The anomaly which size is small in training dataset is more difficult to detect when compared to the normal or other kind of anomaly. As for the muti-anomaly detection, it is also a classical challenge. However, SS-VTCN achieves excellent performance at 40% labeled data.

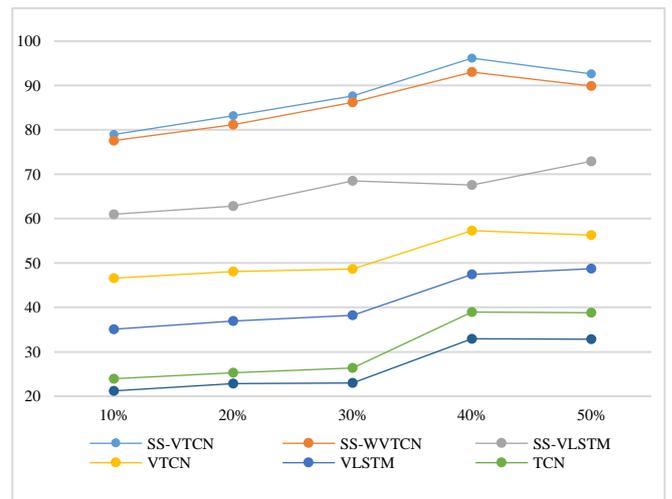

Figure 6. The comparison of $Avg$ performance

The network architecture of SS-VTCN is innovative, which synchronously achieve time series detection and get the latent representations. To search a suitable rate of labeled data, we set five layers for the experiment and use the average

$F1$ score of each class named $Avg$. As shown in Figure 6, all models achieve the relative best performance for DS2OS-B at 40% labeled data. It means that the most suitable rate of labeled data is 40%, which the cost of labeling is not huge. In addition, SS-VTCN almost achieve the best performance at each of rate of labeled data. Moreover, SS-VTCN achieve the balanced and excellent performance of the normal and other different class anomaly.

In conclusion, there is no doubt that SS-VTCN is the best model for solving the muti-anomaly detection problem in IoT communication. Besides, the adoption of unlabeled data brings obvious performance improvement. Actually, the semi-supervised models show natural advantage for muti-anomaly detection compared with supervised models. Moreover, the specific rectification principles have obvious effective influence during the training process. The experiment results indicate that the semi-supervised model SS-VTCN with the rectification principles can better utilize the advantages of VAE and TCN. Meanwhile, SS-VTCN can effectively improve the performance of muti-anomaly detection in IoT communication.

Table 1. Comparison between semi-supervised models at DS2OS-A by averaging across 10 independent runs

|  | Class | SS-VTCN | SS-WVTCN | SS-VLSTM | VTCN | VLSTM | TCN | LSTM |
|---|---|---|---|---|---|---|---|---|
| 10% | 0 | 97.04 | 96.47 | **98.69** | 88.36 | 64.34 | 47.62 | 43.20 |
|  | 1 | **98.91** | 98.32 | 42.12 | 32.47 | 26.82 | 17.58 | 15.83 |
|  | 2 | **64.60** | 62.33 | 60.15 | 42.13 | 28.98 | 19.34 | 16.17 |
|  | 3 | **55.30** | 53.27 | 43.02 | 23.42 | 20.26 | 11.26 | 9.62 |
|  | Avg | **78.96** | 77.60 | 61.00 | 46.60 | 35.10 | 23.95 | 21.21 |
| 20% | 0 | **98.96** | 96.34 | 96.01 | 89.67 | 66.31 | 49.61 | 46.92 |
|  | 1 | **87.08** | 85.79 | 61.38 | 35.18 | 28.12 | 19.27 | 13.53 |
|  | 2 | **63.83** | 62.16 | 56.37 | 44.21 | 30.04 | 20.04 | 18.67 |
|  | 3 | **82.92** | 80.49 | 37.63 | 23.16 | 23.26 | 12.36 | 12.12 |
|  | Avg | **83.20** | 81.20 | 62.85 | 48.06 | 36.93 | 25.32 | 22.81 |
| 30% | 0 | **98.42** | 97.22 | 97.79 | 87.26 | 69.14 | 51.11 | 47.02 |
|  | 1 | **97.69** | 97.34 | 84.02 | 34.79 | 29.07 | 18.23 | 13.98 |
|  | 2 | **79.69** | 77.18 | 61.98 | 48.37 | 32.61 | 20.92 | 19.46 |
|  | 3 | **74.72** | 72.96 | 30.15 | 24.13 | 22.12 | 15.34 | 11.41 |
|  | Avg | **87.63** | 86.18 | 68.49 | 48.64 | 38.24 | 26.40 | 22.97 |
| 40% | 0 | *99.26* | 97.82 | 95.76 | 92.24 | 80.44 | 73.94 | 66.84 |
|  | 1 | *99.62* | 95.46 | 86.01 | 63.32 | 47.53 | 40.42 | 31.72 |
|  | 2 | *95.22* | 91.28 | 57.55 | 47.17 | 36.38 | 23.19 | 18.42 |
|  | 3 | *90.54* | 87.53 | 30.88 | 26.52 | 25.47 | 18.32 | 14.81 |
|  | Avg | *96.16* | 93.02 | 67.55 | 57.31 | 47.46 | 38.97 | 32.95 |
| 50% | 0 | 98.32 | 96.14 | 85.76 | 90.32 | 81.38 | 74.51 | 69.16 |
|  | 1 | 99.71 | 96.83 | 88.19 | 62.72 | 47.92 | 34.87 | 27.43 |
|  | 2 | 81.43 | 80.16 | 65.28 | 49.17 | 37.16 | 26.08 | 18.92 |
|  | 3 | *90.93* | 86.41 | 52.52 | 22.83 | 28.49 | 19.76 | 15.82 |
|  | Avg | 92.60 | 89.89 | 72.94 | 56.26 | 48.74 | 38.81 | 32.83 |

## VI. CONCLUSION AND FUTURE WORK

This paper proposes a novel semi-supervised model named SS-VTCN for muti-anomaly detection in IoT communication. The adoption of rectification principles authentically boosts the performance of detection. Moreover, the experimental results prove the effectiveness of our proposed model. However, the performance of small size anomaly detection can be improved. For future studies, we plan to test in larger size dataset and improve SS-VTCN to solve more complex problems.